\documentclass[aps,prl,superscriptaddress,reprint]{revtex4-1}
 \usepackage{amsmath} \usepackage{amsfonts}
\usepackage{amssymb} 
\usepackage{array}
\usepackage{graphicx} 
\usepackage{verbatim}

\begin{document}

\title{Competition of Pairing and Nematicity in the
Two-Dimensional Electron Gas}

\author{Katherine A. Schreiber}
\affiliation{Los Alamos National Laboratory, Mail Stop E536, Los Alamos, NM 87545}
\author{G\'{a}bor A. Cs\'{a}thy}
\affiliation{Department of Physics and Astronomy, Purdue University, West Lafayette, IN 47907; email: gcsathy@purdue.edu}
\affiliation{Birck Nanotechnology Center, Purdue University, West Lafayette, IN 47907}

\date{\today}

%Abstract
\begin{abstract} %  Abstract text, approximately 150 words. 

Due to its extremely rich phase diagram,
the two-dimensional electron gas exposed to perpendicular magnetic field
has been the subject of intense and sustained study.
One particularly interesting problem in this system is that of the half-filled Landau level, where
the Fermi sea of composite fermions, a fractional quantum Hall state arising from
a pairing instability of the composite fermions, and the quantum Hall nematic were observed
in the half-filled $N=0$, $N=1$, and $N \geq 2$ Landau levels, respectively.
Thus different ground states developed in different half-filled Landau levels.
This situation has recently changed, when evidence for both the paired fractional quantum Hall state
and the quantum Hall nematic was reported in the half-filled $N=1$ Landau level.
Furthermore, a direct quantum phase transition between these two ordered states was found.
These results highlight an intimate connection between
pairing and nematicity, a topic of current interest in several strongly correlated systems,
in a well-understood and low disorder environment.
\end{abstract}

\maketitle

%Table of Contents
%\tableofcontents

% Heading 1
\section{INTRODUCTION}

The interplay of pairing and charge order is of current interest in
a growing number of strongly correlated
electron systems. It has long been suspected that in 
NbSe$_2$ there is a connection between 
superconductive and charge density wave phases \cite{nbse1,nbse2,nbse3}.
These two phases survive in 2$H$-NbSe$_2$ in the single layer limit \cite{nbse4}
and develop in other transition metal dichalcogenides, such as TaS$_2$ \cite{tas2,tas3,tas4}
and TiSe$_2$ \cite{tise1,tise2}.
Most recently the interplay of pairing and charge order has been intensely studied
in high temperature superconductors, both in cuprates \cite{high1,high2,high3} 
and in iron pnictides \cite{iron1,iron3}.
The coupling of superconductive and nematic orders 
is thought to play a role in understanding unconventional superconductivity and it may
enable tuning superconductivity through manipulating the nematic \cite{high4,high5,fernandesAr,proustAr}.

Paired and nematic phases are also known to form in the two-dimensional
electron gas (2DEG) confined to GaAs/AlGaAs heterostructures exposed to perpendicular magnetic fields.
This system supports an astonishingly large number of phases. 
Of these phases, fractional quantum Hall states (FQHS) at even denominator filling factors \cite{willett,jim90,pan99}
are thought to be due to $p$-wave pairing of composite fermions \cite{mooreRead,greiter}, 
the emergent particles of the fractional quantum Hall regime \cite{jain}.
Prominent paired FQHSs form in the $N=1$ Landau level.
A different phase with charge order forms in high Landau levels with $N \geq 2$
\cite{fogler,moessner,kivel,fradkin,lilly99,du99}. This phase is called
the quantum Hall nematic (QHN) or the stripe phase.
While the paired FQHSs and QHN were known for more than 20 years, they developed
in different Landau levels and hance a transition between them did not seem possible.
It has only recently been learned that the QHN may also be stabilized in the $N=1$
Landau level and that a direct phase transition may be induced
between the paired FQHSs and the QHN \cite{pan14,kate1,kate2,kate3}. 
These discoveries offered the chance to study the interplay of pairing and nematicity 
in the 2DEG.

It is generally understood that both pairing and nematicity are
driven by a particular type of effective interaction between the electrons,
specifically an interaction that consists of a short range attractive and a long range repulsive part.
Such interactions are also realized in the 2DEG. However, paired and nematic phases of the 2DEG
contrast those in unconventional superconductors in several ways. 
First, as already mentioned, pairing
of the composite fermions is expected to be $p$-wave in nature \cite{mooreRead,greiter}. 
In contrast to superconductors, in the 2DEG pairing of the composite fermions
is driven by electron-electron interactions, rather than electron-phonon interactions. 
Second, pairing of composite fermions in the fractional quantum Hall regime occurs
in the presence of edge states, highlighting therefore the importance of topological aspects.
Finally, the interplay of pairing and nematicity in the 2DEG
occurs under certain desirable conditions: 
the 2DEG is a well-understood and low disorder system,
the physics of the 2DEG is a single band physics, and
the spin is weakly coupled to the orbital degree of freedom.

The scope of this article is limited to electrical transport investigations of
the half-filled $N=1$ Landau level of the 2DEG confined to GaAs/AlGaAs
hosts which are exposed to a perpendicular magnetic field.
Much of the data discussed was acquired in samples at high
hydrostatic pressure, leading to a phase transition between paired
FQHSs and the QHN.
There are other 2DEGs that exhibit FQHSs associated with pairing of composite
fermions, such as the ones confined to ZnO/MgZnO  heterostructures \cite{zno}, 
bilayer graphene \cite{gr0,gr1,gr2}, and monolayer graphene \cite{gr3,gr4}.
However, in contrast to GaAs/AlGaAs, in these hosts 
the QHN so far has not been observed and will not be  further discussed.

\section{SNAPSHOTS OF PHASES OF THE TWO-DIMENSIONAL ELECTRON GAS}

The single particle energy spectrum of a 2DEG placed in a perpendicular magnetic
field consists of a set of discreet and degenerate Landau levels  \cite{jainBook,ezawaBook}. In the absence of the
valley degree of freedom, Landau levels of the 2DEG confined to GaAs/AlGaAs
heterostructures are labeled by the orbital index $N=0,1,2,...$ and spin.
At $B=1$~T, the energy scales associated with the orbital and spin degree of
freedom of non-interacting electrons on GaAs are the cyclotron energy $\hbar \omega_c =22$~K and 
the Zeeman energy $E_z=0.3$~K, respectively. 
It is customary to call the number of filled Landau levels the Landau level
filling factor $\nu= hn/eB$, where $n$ is the electron areal density, $h$ the Planck
constant, $e$ the charge of the electron, and $B$ the magnetic field. 
Because of the spin degree of freedom, in the GaAs/AlGaAs system
each orbital energy level is split into two spin branches at experimentally relevant magnetic fields.
The $N=1$ Landau level, also called the second orbital Landau level, 
corresponds to the $2 < \nu <4$ range of filling factors;
of this range the $2 < \nu <3$ is the lower spin branch, whereas the $3 < \nu <4$ range is
the upper spin branch.

The 2DEG supports a treasure trove of phases that fall into one of the
two distinct classes: topological phases and traditional Landau phases.
Topological phases are insulating in their bulk and carry current along their edges,
are degenerate, exhibit an energy gap in their excitation spectrum, and in some cases may have exotic
quasiparticle excitations. Because of the presence of an energy gap, these phases are 
incompressible. Integer quantum Hall states (IQHSs) forming at $\nu=p$ 
and fractional quantum Hall states (FQHSs) forming at  $\nu=p/(2p \pm 1)$ and $\nu=p/(4p \pm 1)$,
with $p$ a positive integer, are examples of topological phases in the 2DEG \cite{klitz,tsui1,jainAr}. 
Their magnetoresistive signatures are a vanishing longitudinal magnetoresistance $R_{xx}=0$
and a quantized Hall resistance $R_{xy}=h/fe^2$ measured at the Landau level filling factor $\nu=f$,
where $f$ is either an integer or a fraction.

In contrast to IQHSs, FQHSs are driven by the Coulomb interaction between the electrons.
However, the interaction energy in not a small parameter and thus 
FQHSs cannot be adequately described by a Landau type Fermi liquid theory \cite{jainBook}.
In fact in the fractional quantum Hall regime the Coulomb interaction becomes the dominant energy scale.
Because of this, the 2DEG is said to be a strongly correlated system.
In order to make progress with the seemingly intractable many-electron correlations, 
Jain proposed a canonical transformation of the system into a set a weakly interacting
emergent particles called composite fermions \cite{jain}.
A composite fermion is an electron with an even number of quantized votices attached to it.
The most common composite fermion is the one with two vortices.
Since the number of magnetic flux tubes piercing a 2DEG is the same as the number
of vortices, an alternative formulation is an electron
binding to an even number of fictitious magnetic flux quanta. As a result of the flux attachment procedure,
the composite fermions experience an effective magnetic field that is significantly
reduced from the value of the externally applied magnetic field. 
The composite fermion theory naturally 
accounts for a large number of FQHSs as IQHSs of the composite
fermions \cite{jain}.

A collection of composite fermions describes the factional quantum Hall regime remarkably well.
It is, however, important to appreciate that composite fermion framework is not just simply a useful
mathematical mapping. Indeed, there are numerous
experimental results attesting the formation of the composite fermions. Examples are 
energy gap scaling \cite{gap1},   
surface acoustic wave propagation experiments \cite{saw0},
geometric resonance measurements \cite{geom1}, and magnetic focusing \cite{focus1}
in the fractional quantum Hall regime.

\subsection{Half-filled $N=0$ Landau Level}

Because of a lifted spin degeneracy, in the GaAs/AlGaAs system there two energy levels with
$N=0$; these levels are half-filled at $\nu=1/2$ and $\nu=3/2$. 
Early experiments reported a large number of FQHSs in the $N=0$ Landau level.
However, a FQHS was conspicuously missing at $\nu=1/2$ and $\nu=3/2$ \cite{willett,gap1}.
Observations at these two filling factors indicated a gappless, compressible state instead.
Halperin, Lee, and Read have investigated this state and found 
a Fermi sea of composite fermions \cite{hlr}. Indeed, at these two filling factors a 
canonical transformation maps the electron system exposed to a strong magnetic field  
into a set of composite fermions at zero effective magnetic field.
The effective mass of the composite fermions is not a free parameter of the theory,
but it encodes the effect of the electron-electron interaction.

Most recently, the Fermi sea of composite fermions in the $N=0$ Landau level was reexamined in the limit
of exact particle-hole symmetry \cite{son1}. According to this analysis, the composite fermions
at $\nu=1/2$ in a system of non-relativistic electrons must be Dirac particles. 
Theories constrained by particle-hole symmetry
naturally account for a Fermi sea at $\nu=1/2$ \cite{son1,son2,son3}. 
However, in realistic 2DEGs particle-hole symmetry is 
broken because of significant Landau level mixing and of finite width effects and therefore
it is not clear whether or not these theories strictly apply.
To conclude, at $\nu=1/2$ and $\nu=3/2$ of the $N=0$ Landau level there is
a Fermi sea of composite fermions. Certain
aspects of this Fermi sea are, however, still under active investigation \cite{son4}.

\begin{figure*}[t]  %h
\includegraphics[width=5in]{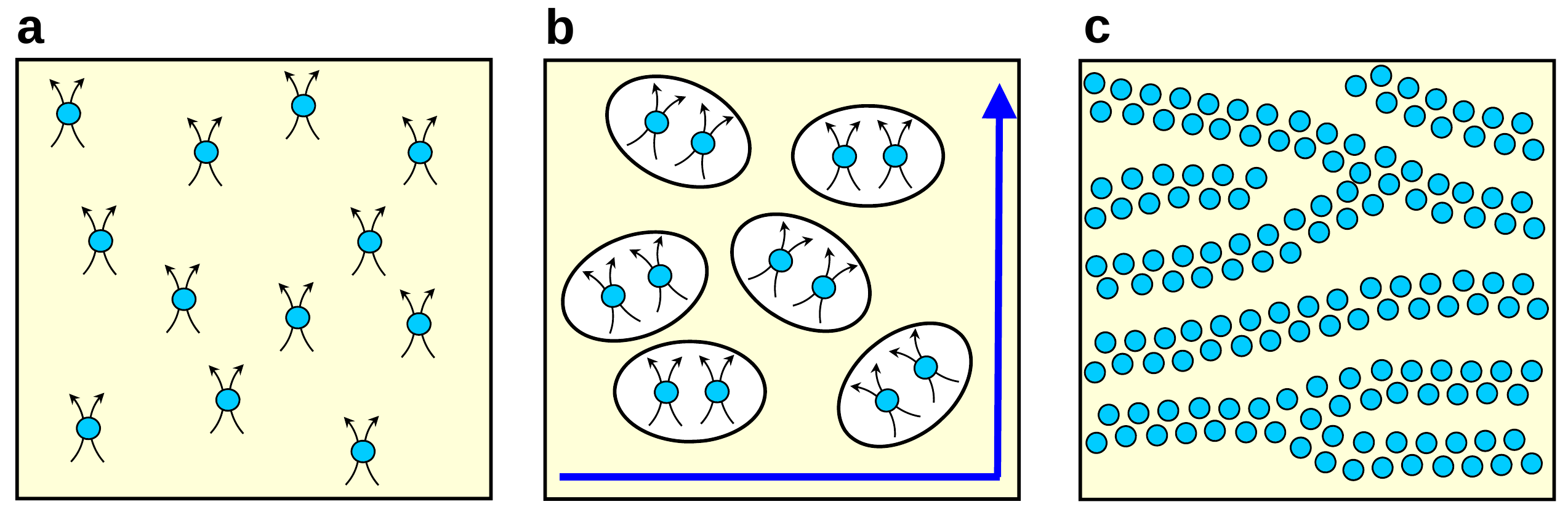}
\caption{A representation of phases in half-filled Landau levels.
(\textbf{a}) Fermi liquid of flux-two composite fermions.
Composite fermions are depicted as electrons with two quantized fluxlines attached.
\newline
(\textbf{b}) Paired FQHSs consist of Cooper pairs of composite fermions and
possess edge states. 
\newline
(\textbf{c}) The QHN is a
filamentary electronic phase that breaks rotational symmetry.
Adapted from Schreiber et al. \cite{kate3}.
}
\label{fig1}
\end{figure*}

\subsection{Half-filled $N=1$ Landau Level}

Half-filled energy levels at $N=1$ occur in the GaAs/AlGaAs system at $\nu=5/2$ and $\nu=7/2$.
In contrast to results from $N=0$, in the half-filled $N=1$ Landau level 
FQHSs develop both at $\nu=5/2$ \cite{willett,jim90,pan99} and $\nu=7/2$ \cite{eisen}.
Owing to the even denominator of the filling factor, it
was realized early on that these two FQHSs do not belong to the sequence of FQHSs
predicted by the model of non-interacting composite fermions. It is said that the even denominator
FQHSs forming at $\nu=5/2$ and $\nu=7/2$ are topologically distinct from the odd denominator
FQHSs forming in the $N=0$ Landau level.

Composite fermions form not only in the $N=0$ Landau level, but also in the $N=1$
Landau level \cite{cf1,cf2}. However, the theory does not guarantee that the composite fermions will be
weakly interacting. It turns out that in the $N=1$ Landau level
the effective interaction between the composite 
fermions is attractive and this attraction is sufficiently
strong to induce a pairing instability of the Fermi sea of composite fermions
\cite{mooreRead,greiter,pair2,pair3,pair4,pair5,pair6}, 
opening therefore an experimentally measurable energy gap.
The Pfaffian proposed by Moore and Read is a candidate wavefunction for such a pairing instability
of the Fermi sea \cite{mooreRead}. Because of the large magnetic fields
these FQHSs form at, Cooper-like pairing of the composte fermions neccessarily has to involve
aligned spins, hence the pairing is $p$-wave in nature.
The Pfaffian originally proposed for these states indeed maps into a $p_x +ip_y $ superconductor \cite{greiter}.
It is then customary to call the $\nu=5/2$ and $\nu=7/2$ FQHSs paired.

Besides the Pfaffian description,
the FQHSs at $\nu=5/2$ admits other competing descriptions that are distinct from the Pfaffian.
Examples are the anti-Pfaffian \cite{anti1,anti2}, the (3,3,1) Abelian state \cite{halperin331}, 
a variational wave function based on an anti-symmetrized bilayer state \cite{anti3}, 
the particle-hole symmetric Pfaffian \cite{son1,dirac2}, a stripe-like alternation of the Pfaffian 
and anti-Pfaffian \cite{anti4}, and other exotic states \cite{wen90,wen91}. 
%Pairing of the composite fermions is thus compatible with different topological orders.
Numerical studies place the $\nu=5/2$ FQHS
in the Pfaffian universality class,  ensuring the paired nature of the FQHS
\cite{sim0,sim1,sim2,sim3,sim4,sim5,sim6,sim7,sim8,sim9}.
Results of ongoing experimental investigations do not yet converge on the nature of
the topological order of this state
% have not yet provided a converging picture on the topological order 
\cite{exp0,exp1,exp2,exp3,exp4,exp5,exp6}.

\subsection{Half-filled $N \geq 2$ High Landau Levels}

The 2DEG also supports various types of charge density waves that belong to the family of
traditional Landau phases. The most well-known of these is the Wigner
crystal. However, the pioneering Hartree-Fock theory of Koulakov, Fogler and Shklovskii
\cite{fogler} and also of Moessner and Chalker \cite{moessner} predicted more intricate charge order
in high Landau levels, i.e. for $N \geq 2$. Away from half-filling, these theories found
isotropic charge density waves termed electronic bubble phases \cite{fogler,moessner}. 
Furthermore, near half-filling unidirectional charge density waves called stripe phases
were anticipated \cite{fogler,moessner}. Stripe phases were also reported in exact diagonalization \cite{nem1}
and density matrix renormalization group studies \cite{nem2}.
By considering fluctuation effects beyond the 
mean field treatment of the Hartree-Fock approach,  Fradkin and Kivelson
found that a richer set of electronic crystals is allowed that includes the nematic and the smectic
\cite{kivel,fradkin}. Further theory work strengthened the case for these phases \cite{nem3,nem4}.

Anisotropic phases at half-filling discovered by Lilly et al. \cite{lilly99} and Du et al. \cite{du99}
in 1999 in the $N=2$ and $3$ Landau levels of 2DEGs confined to GaAs/AlGaAs, i.e.
at filling factors $\nu=9/2$, $11/2$, $13/2$,..., were associated with stripe phases at their discovery. 
The Hall resistance at these filling factors is not quantized. Anisotropy was also
detected in microwave pinning resonance \cite{micro} and surface acoustic wave propagation 
\cite{saw1,saw2}. Since experiments typically detect broken rotational symmetry,
we still lack information on translational order in these phases \cite{qian}.
Therefore anisotropic phases at $\nu=9/2, 11/2, 13/2,...$ are referred to as
the QHN, or simply the nematic \cite{jimAr}. In contrast to FQHSs, the QHN are compressible.
A rendering of the QHN and other phases at half-filling can be seen in \textbf{Figure \ref{fig1}}.

It is important to note that the nematic is a widely used term in condensed matter physics
for fundamentally distinct types of anisotropic behavior. 
2DEGs in AlAs quantum wells \cite{alas} and the surface states of elemetal Bi \cite{bi} have anisotropic mass,
hole gases in Si doped (311)A interface of GaAs/AlGaAs exhibit anisotropic scattering \cite{scatt},
and under certain conditions fractional quantum Hall states are anisotropic \cite{nFQH1,nFQH2,nFQH3}.
All these examples fall under the umbrella of nematicity, but lack charge order. 
In contrast, the QHN in half-filled Landau levels in the 2DEG 
and nematic phases in strongly correlated materials, such as the cuprates, iron
pnictides, and layered superconductors, possess charge order. In particular, the
QHN at $\nu=9/2$ can be thought of as a phase with an
interpenetrating filaments of $\nu=4$ and $\nu=5$ regions. 
The width of these regions is about three classical cyclotron radii \cite{fogler,friess}.

Within the Hartree-Fock description, at the origin of stripes and related QHN one finds the overlapping
electronic wavefunctions which soften the effective interaction between the electrons \cite{fogler,moessner}.
The spontaneous formation of stripes can also be accounted for by a
Pomeranchuk instability of the Fermi sea that occurs
when the Fermi liquid parameter in the $l=2$ angular momentum  channel is less than $-1$.
Recent variational Monte Carlo calculations in high Landau levels found numerical evidence for
a Pomeranchuk instability \cite{th4}.

\section{PROXIMITY OF PAIRED FRACTIONAL QUANTUM HALL STATES TO NEMATIC PHASES}

In a large number of experiments performed on 2DEGs in strictly perpendicular
magnetic fields, an isotropic FQHS was reported at $\nu=5/2$ and $\nu=7/2$.
In contrast, experiments in tilted magnetic fields, i.e. with a non-zero
in-plane magnetic field, anisotropy develops at both $\nu=5/2$ and $\nu=7/2$. 
The nature of anisotropic phases under tilt is nuanced. Refs. \cite{tilt1,tilt2,tilt3} report
a compressible nematic phase under tilt.
One experiment found an anisotropic FQHS, an incompressible state, at small and moderate tilt angles \cite{nFQH2}.
At extreme tilt angles an isotropic compressibe phase
reminiscent of the Fermi sea of composite fermions is recovered \cite{tilt3}.
These experiments show that the isotropic FQHSs in the half-filled $N=1$ Landau level
observed in perpendicular magnetic fields are energetically close to a nematic phase \cite{jimRev}.

\begin{figure*}[t]  %h
\includegraphics[width=5in]{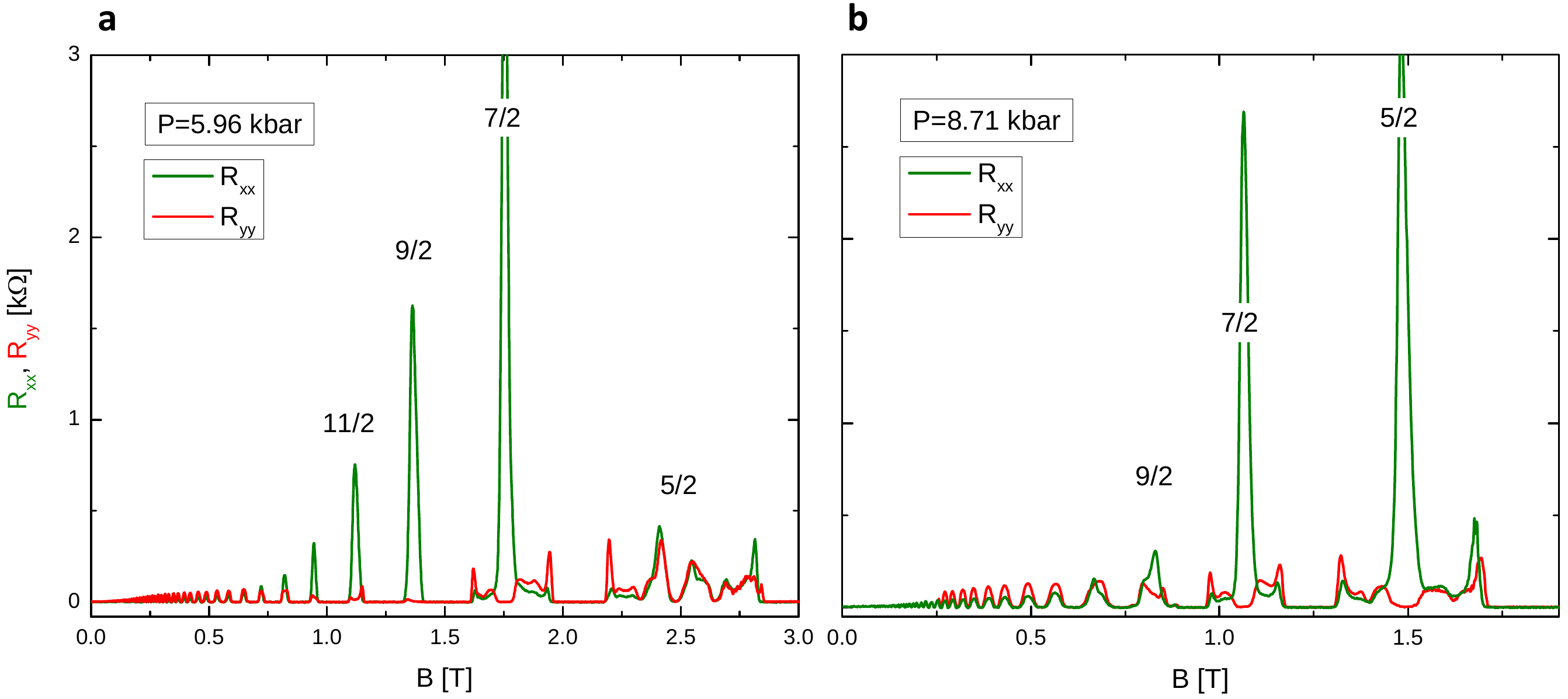}
\caption{Magnetoresistance traces measured near
$T \simeq 12$~mK along two mutually perpendicular directions
of the GaAs crystal. Data was acquired at
 $P=5.96$~kbar (panel  (\textbf{a})) and at $P=8.71$~kbar (panel (\textbf{b})).
Based on data from Samkharadze et al. \cite{kate1} and Schreiber et al. \cite{kate2}.
}
\label{fig2}
\end{figure*}

The development of anisotropy in a 2DEG under tilt is not surprising since an in-plane
component of the magnetic field couples to the nematic order parameter and it therefore
favors nematicity \cite{tilt-th1,tilt-th2}. 
To illustrate this, it is useful to consider the analogy between
between the 2DEG and a system of non-interacting spins.
In the latter system, an external magnetic field 
couples to the magnetic order and it induces a finite magnetization.
Similarly, an in-plane magnetic field induces nematicity in the 2DEG.
By the same token, uniaxial strain also favors nematicity
\cite{sunandra}.

The analogy between the 2DEG and a spin system may also be extended to the 
case of spontaneous symmetry breaking.
In a system of interacting spins, spontaneous magnetization or ferromagetism
develops in the absence of any externally applied fields.
Similarly, in a 2DEG the QHN develops in the absence of any
externally applied symmetry breaking fields.
We note that the magnetized phase of a paramagnet placed in an external magnetic
field is not identical to the spontaneous ferromagnetic phase; for example, 
the two phases do not share the same correlation functions. By analogy,
the compressible nematic phase in tilted magnetic fields at $\nu=5/2$
is likely related to, but it is not necessarily identical to
the QHN spontaneously forming at $\nu=9/2$.
The exact nature of the relationship between these two nematic phases
is yet to be determined.

The QHN at $\nu=9/2$ develops in the absence of any
externally applied symmetry breaking fields favoring nematicity. 
However, it is widely recognized that there is an internal or native symmetry breaking
field present in the GaAs/AlGaAs samples \cite{lilly99,du99,jimRev}. 
This internal field is responsible for locking the direction of the nematic filaments
with one crystallographic axis of the GaAs. In many cases the maximum of the magnetoresistance $R_{xx}$
is in the [1$\bar{1}$0] crystal direction, while the vanishing magnetoresistance $R_{yy}$ is measured
along the [110] direction \cite{lilly99,du99}. The magnitude of this internal field
is known to be small; the associated potential is estimated a few mK per electron \cite{jimRev,mk1}. 
Remarkably, the nature of this internal symmetry breaking field has not yet been identified and thus remains one
of the important outstanding questions for the QHN. Recent results on this topic
can be found in Refs.\cite{dir1,dir2,dir3,dir4}. The internal field
that aligns the nematic domains is analogous to the magnetic anisotropy in a crystalline
ferromagnet. Ferromagnetic domains exist in the absence of magnetic anisotropy, but may
point in random directions. An anisotropic interaction with the crystal lattice
orients the ferromagnetic domains.
% resulting in a large macroscopic magnetization and a nematicity of the Ising-type \cite{sayan}.
Similarly, the internal field orients the nematic domains, leading
to a dramatic resistance anisotropy not seen in nematic phases of other strongly correlated materials.

We now turn our attention to the question of proximity to a nematic in the absence
of an external symmetry breaking field. Early theory work of Rezayi and Haldane found that
by tuning the effective electron-electron interaction away from its Coulomb expression,
a phase transition from a paired FQHS to the stripe phase occurs at $\nu=5/2$ \cite{haldane-str}.
However, it was not known whether the interaction potential 
required for such a phase transition may be achieved in the experimentally accessible parameter space.
Subsequent numerical work noted regions of poor overlap 
of the numerical wavefunction and the Pfaffian \cite{sim0,sim1,sim2,sim3,sim4,sim5,sim6,sim7}. 
However, in lack of experimental observations of the QHN at $\nu=5/2$,
in these regions of poor overlap the QHN or the related stripe phase 
was often not considered \cite{sim0,sim1,sim2,sim3,sim4}.

A further indication for the proximity to a nematic instability in the half-filled $N=1$ Landau level 
came from a recent experiment performed in a purely perpendicular magnetic field \cite{pan14}. 
In a sample of low density $n=5.0 \times 10^{10}$~cm$^{-2}$, Pan et al. found an incipient 
anisotropy at $\nu=7/2$, with a resistance anisotropy ratio of 2. At $\nu=5/2$ an isotropic FQHS
was observed \cite{pan14}.

\section{HIGH PRESSURE STUDIES OF THE $N=1$ LANDAU LEVEL}

We have seen that the paired FQHS at $\nu=5/2$ is close to a nematic phase.
However, in the 30 year history of experimental work at $\nu=5/2$, 
anisotropic behavior at this filling factor was never observed in the absence of 
a symmetry breaking field favoring nematicity
\cite{exp0,exp1,exp2,exp3,exp4,exp5,exp6,xia04,choi08,pan08,kumar10,shay10,shay11,deng12,gamez,gaas3,gerv,
dean08,nodar11,pan12,nuebler10,watson15,shi15,nodar17}. 
This state of affairs changed recently with the
discovery of the QHN at $\nu=5/2$ 
in 2DEGs under high hydrostatic pressure \cite{kate1,kate2,kate3}. 

The application of hydrostatic pressure is a wide-spread technique in condensed matter physics.
This is because high pressure impacts electronic bands by
changing the lattice constant and therefore the Bloch wavefunction.
The largest pressures are achieved in diamond anvil cells. However, the small volume available
makes this technique extremely challenging for GaAs transport measurements. 
Pressure clamp cells afford a reasonably large working volume while generating sufficiently high pressures, exceeding $20$~kbar.

Parameters under high pressures of 2DEGs confined to GaAs structures 
are well documented \cite{p1,p2}. Perhaps the most striking effect 
is the change of band energies and of dopant energy levels with pressure. As a result,
and increasing pressure causes a decrease in the areal density of the electrons \cite{kate1,p1}. 
Lower densities are necessarily accompanied by reduced electron 
mobilities. In addition, under quasi-hydrostatic conditions,
the mobility may be further reduced by scattering due to small fluctuations of the density caused by minute
pressure variations of the frozen pressure transmitting fluid along the plane of the 2DEG. 

In contrast to in-plane magnetic fields and uniaxial strain, hydrostatic pressure is 
a tuning parameter that does not explicitly break the rotational symmetry in the plane of the 2DEG. 
Indeed, hydrostatic pressure shrinks the unit cell of the GaAs crystal without 
causing a deformation favoring a particular crystal direction.

\begin{figure*}[t]  %h
\includegraphics[width=5in]{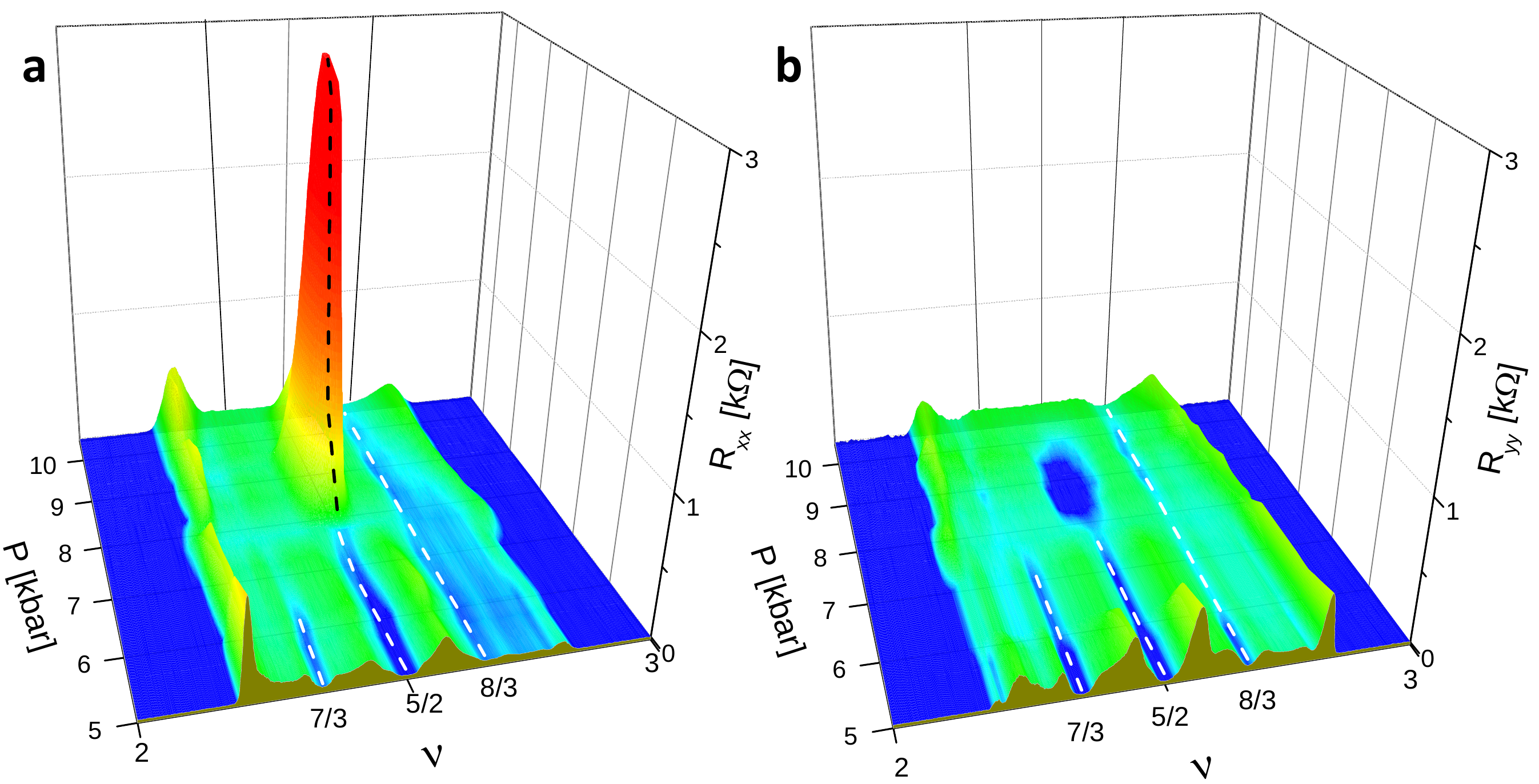}
\caption{
Magnetoresistance manifolds $R_{xx}$ (panel (\textbf{a})) and $R_{yy}$ (panel (\textbf{b}))
in the lower spin branch of the $N=1$ Landau level, as measured at $T \simeq 12$~mK.
At $\nu=5/2$ we observe a paired FQHS at $P<7.8$~kbar, the QHN at $7.8 <P<10$~kbar, 
and an isotropic Fermi liquid at $P>10$~kbar. 
Based on data from Samkharadze et al. \cite{kate1}.
}
\label{fig3}
\end{figure*}

\subsection{The observation of the quantum Hall nematic at $\nu=5/2$ and $\nu=7/2$}

The lower spin branch of the $N=1$ Landau level in a 2DEG in a perpendicular magnetic field
was studied at high hydrostatic pressures in Ref.\cite{kate1}.
The sample had an electron density $n=2.8 \times 10^{11}$~cm$^{-2}$, 
mobility $\mu = 15 \times 10^6$~Vs/cm$^2$, and its
structure was based on a $30$~nm symmetrically doped quantum well 
with a short-period superlattice doping scheme.
At $P=5.96$~kbar an isotropic FQHS at $\nu=5/2$ was found, as evident
by the vanishing magnetoresistance shown in \textbf{Figure \ref{fig2}a}
and a quantized Hall resistance (not shown).
This FQHS is adiabatically connected to the one in the ambient. 
An increase in pressure, however, leads to a qualitatively
different magnetoresistance near $\nu=5/2$.
Indeed,  magnetoresistance traces at $P=8.71$~kbar measured along mutually perpendicular
crystal axes of th GaAs exhibit a dramatic anisotropy at $\nu=5/2$ \cite{kate1}.
This is shown in \textbf{Figure \ref{fig2}b}.
The observation of anisotropy at $P=8.71$~kbar indicates
a ground state at $\nu=5/2$ that breaks rotational symmetry. 

A comparison of the anisotropic magnetortesistance at $\nu=5/2$ measured at $P=8.71$~kbar \cite{kate1} and that
at $\nu=9/2$ measured in the ambient \cite{lilly99,du99} indicates that the associated two ground states are similar.
First, both resistance anisotropies develop spontaneously, i.e. in the absence of
any externally applied symmetry breaking fields favoring nematicity.
Second, the temperature dependence of $R_{xx}$ and $R_{yy}$ is very similar,
i.e. exponentially diverging one from another in both cases \cite{lilly99,du99}.
Third, in both cases, the anisotropy ratio $R_{xx}/R_{yy}$ at the lowest temperatures 
is vey large, exceeding $100$. Finally, in both cases anisotropy develops
over a limited span of filling factors $\Delta \nu \simeq 0.15$
centered to half-integer filling factors \cite{kate1}. In contrast, the
resistance anisotropy induced by an external in-plane magnetic field
at $\nu=5/2$ is present over a considerably wider range of filling factors $\Delta \nu \simeq 0.6$ \cite{tilt1,tilt2,tilt3}. 
We thus conclude that the anisotropic ground state at $\nu=5/2$ and at $P=8.71$~kbar,
shown in  \textbf{Figure \ref{fig2}b}, is a genuine QHN.

While hydrostatic pressure is not expected to break the rotational symmetry
in the plane of the 2DEG, an undesired tilt of the 2DEG inside the pressure cell
will have such an effect.
This situation may occur if one corner of the sample is grabbed by the
teflon lining of the pressure cell during pressurization. 
In a different experiment we found that the sample did indeed tilt \cite{kate4}.
When this situation was
encountered, the measured $n$ at the particular pressure  $P$ 
no longer followed the expected linear dependence of the density on pressure
measured in the absence of tilt. In addition, in contrast
to the narrow range anisotropy $\Delta \nu \simeq 0.15$ seen
in  \textbf{Figure \ref{fig2}}, the accidentally tilted sample
was anisotropic over a significantly larger range of filling factors.
Such results constitute evidence for the absence of any significant symmetry
breaking in-plane magnetic field for data shown in \textbf{Figure \ref{fig2}}.

A strong QHN develops not only at $\nu=5/2$, but also at $7/2$.
This can be seen for $P=5.96$~kbar and $P=8.71$~kbar 
in \textbf{Figure \ref{fig2}a} and \textbf{Figure \ref{fig2}b},
respectively. However, in contrast to $\nu=5/2$ and $7/2$, 
at $\nu=3/2$  we did not observe the QHN \cite{kate3,kate4}.
We therefore conclude that the critical orbital number for the
QHN, i.e. the lowest orbital quantum number for nematicity, is $N=1$.

The QHN at $\nu=9/2$ in experiments in the ambient and also
at $\nu=5/2$ under high pressures develops in the absence of any
externally applied symmetry breaking fields favoring nematicity. 
However, a small internal field is present that assures locking of the nematic order to the
crystal axes of the GaAs \cite{lilly99,du99,jimRev}. 
Data in \textbf{Figure \ref{fig2}a} show that the orientation of the QHN at $\nu=7/2$
and $9/2$ is the same. Similarly, data in \textbf{Figure \ref{fig2}b} show that
the orientation of the QHN at $\nu=5/2$ and $7/2$ relative to the GaAs crystal axes is the same.
Thus data from Refs.\cite{kate1,kate2,kate3} show that, within a given sample,
the direction of QHN at high pressure is the same as that at ambient pressure.
Pressure does not appear to change the orientation of the QHN with respect to the
GaAs crystal axes.

\subsection{The transition from the paired fractional quantum Hall state to the quantum Hall nematic}

So far we established that there are two ordered phases in the half-filled $N=1$ Landau level: 
a paired FQHS in the ambient and the QHN at some high value of the pressure. 
These two phases are fundamentally different: paired FQHSs are topological phases that
most likely support exotic non-Abelian excitations, while the QHN is a 
traditional Landau phase with charge order.

In the $T=0$ limit, there are two possible arrangements
of these phases along the pressure axis: 1.) the two phases may be contiguous
to each other, with a direct quantum phase transition between them at a critical pressure and
2.) there may be another phase intercalated between them, such as an isotropic
Fermi liquid. In the latter scenario, the $T=0$ phase diagram would have two
critical pressures, one separating the paired FQHS and the Fermi liquid and another
separating the Fermi liquid and the QHN.
Because of the finite base temperature of the refrigerator, these two
arrangements may not be readily distinguished.
However, as discussed below, the first scenario offers the simplest and most elegant interpretation
of the existing high pressure data.

An argument in favor of a direct $T=0$ quantum phase transition 
at $\nu=5/2$,  from a paired FQHS to the QHN, was
first presented in Ref.\cite{kate1}. In this experiment, the change from
 the FQHS to the nematic occured in a narrow pressure range, between
$P = 6.95$ and $P = 8.26$. This sudden change of the phases
suggested a direct phase transition between the two phases. 
Furthermore, interpolated data measured at other pressures was also consistent 
with a direct phase transition. 
A 3D rendering of magnetoresistance data from Ref.\cite{kate1}
plotted against filling factor and pressure, as measured at $T \simeq 12$~mK,
is shown in \textbf{Figure \ref{fig3}}. 
Blue regions in this figure mark vanishing magnetoresistance and dashed lines
are cuts at constant filling factors in the magnetoresistance manifold.
Moving along the $\nu=5/2$ dashed line in \textbf{Figure \ref{fig3}a}, 
a very abrupt transition is seen in $R_{xx}$
near the critical pressure, where $R_{xx}$ rises rapidly.
The transition from the FQHS to the QHN at $\nu=5/2$ is also
seen in \textbf{Figure \ref{fig3}b}: there is a FQHS along the blue trench 
in the low pressure region which
is separated from, but is in near proximity to a wide blue basin centered to $P \simeq 9$~kbar, the
region associated with the QHN.

\begin{figure}[t]  %h
\includegraphics[width=3in]{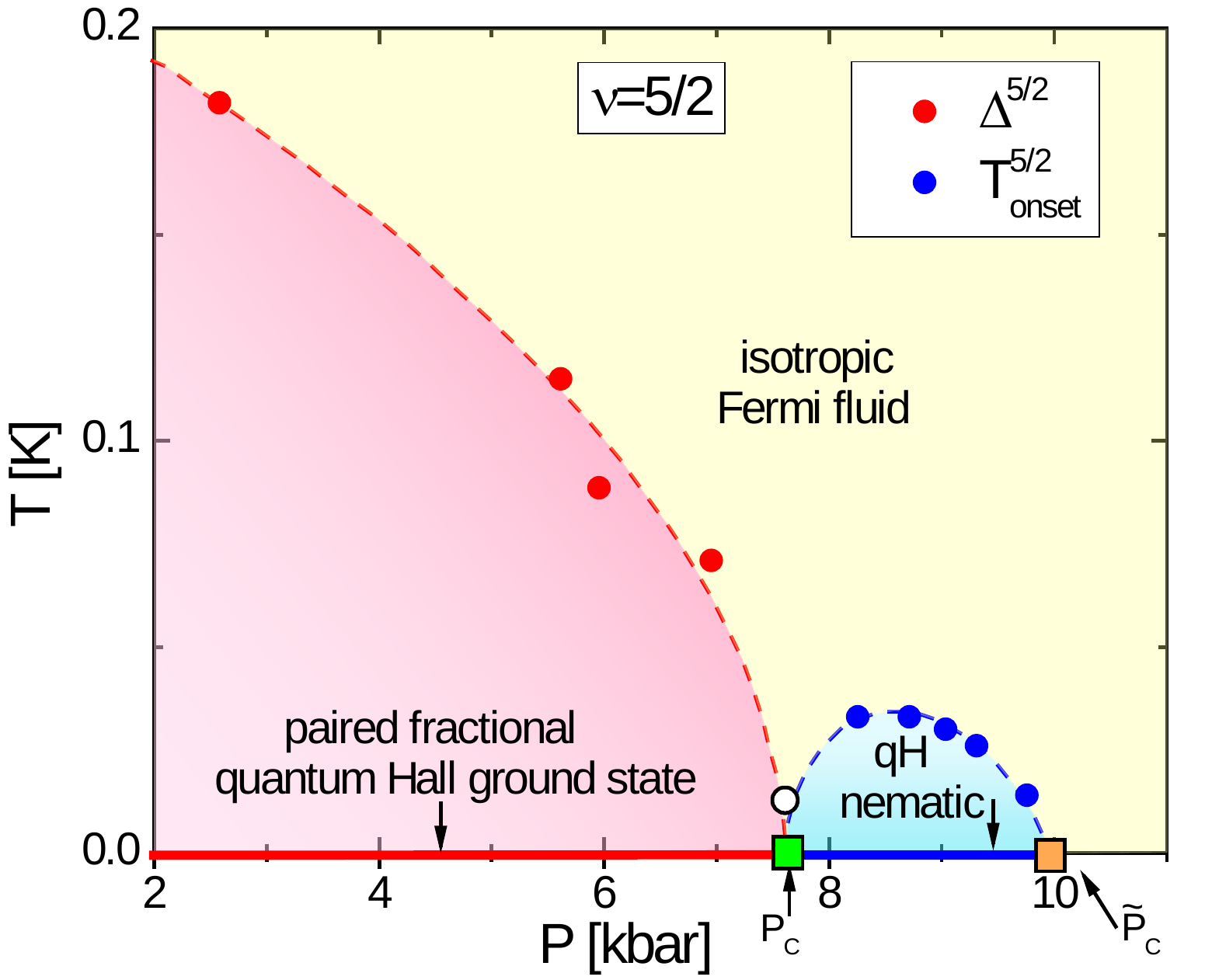}
\caption{
Plot of the energy scales of the ordered phases measured at $\nu=5/2$ 
against the pressure. The open symbol represents an isotropic Fermi liquid state
measured at the lowest temperature accessed $T \simeq 12$. 
In the limit of $T=0$, data are consistent with a direct quantum phase transition from a paired FQHS to
the QHN occurring at the critical pressure $P_c$. At a higher pressure $\tilde{P}_c$ there is 
another quantum phase transition from the QHN to the isotropic Fermi liquid.
 Adapted from Schreiber et al. \cite{kate2}. 
}
\label{fig4}
\end{figure}

Finite temperature studies lent additional weight to a direct $T=0$ phase 
transition from a paired FQHS to the QHN \cite{kate2}.
Each ordered phase at $\nu=5/2$ has an energy scale associated with it:
a FQHS is characterized by an energy gap $\Delta^{5/2}$, while the 
QHN by the onset temperature $T_{onset}^{5/2}$.
$\Delta^{5/2}$ is extracted from an activated temperature behavior, while $T_{onset}^{5/2}$
is estimated by imposing a significant anisotropy $R_{xx} = 2R_{yy}$ 
in a linear interpolation of the measured data. The $P$-$T$ diagram in
\textbf{Figure \ref{fig4}} shows the pressure dependence of these two energy scales
at $\nu=5/2$. We observe that $\Delta^{5/2}$ monotonically decreases
with an increasing pressure. Such a behavior was expected based on decreasing densities
with an increasing pressure. At higher pressures we find that the QHN. 
Dashed lines in \textbf{Figure \ref{fig4}} are guides to the eye.
The extrapolation to $T=0$ of these two energy scales provides the
$T=0$ phase diagram; due to lack of knowledge of the analytical dependence
of the energy scales on pressure and due to the scatter in the data which results in large part
from the pressure changing procedure performed at room temperature, 
an extrapolation to $T=0$ cannot be performed. However, available data is consistent
with a direct quantum phase transition in the limit of $T=0$ 
from a paired FQHS to the QHN. Similar data
was obtained at $\nu=5/2$ in a second sample, labeled $A$ in Ref.\cite{kate3}.
Because in the $P$-$T$ diagram the Fermi liquid is wedged in between the FQHS and
the QHN, in an experiment performed at a non-zero temperature
along a path of increasing pressure one will not observe a direct phase transition, but
a sequence of FQHS, Fermi liquid, and QHN phases.
The open symbol at $P=7.60$~kbar in \textbf{Figure \ref{fig4}} marks such a Fermi liquid
at $T \simeq 12$~mK.

Data obtained at $\nu=7/2$ in sample $A$ of Ref.\cite{kate3} further support
the idea of a direct $T=0$ quantum phase transition 
from a paired FQHS to the QHN. While the QHN also
developed at $\nu=7/2$  in the sample studied in Ref.\cite{kate1}, we did not have the chance to
study this filling factor at low enough pressures at which a FQHS was expected. This is
because repeated thermal cycling to room temperature led to the
explosion of the feedthrough and the destruction of the sample.

In addition to the quantum phase transition from the paired FQHS to the QHN
at $P_c \simeq 8.7$~kbar, in \textbf{Figure \ref{fig4}} there is a second quantum phase
transition at a higher pressure $\tilde{P}_c \simeq 10$~kbar from the QHN to the Fermi liquid.
Becasue of the extremely low electron densities at $\tilde{P}_c$,
we think that this second quantum phase transition is driven by disorder \cite{kate1,kate2,kate3}.

\subsection{The pressure driven transition at half-filling: an example for a transition from a topological to a traditional Landau phase}

So far we argued in favor of a pressure-driven quantum phase transition 
from a paired FQHS to the QHN. 
Quantum phase transitions are ubiquitous in the condensed matter and they
typically occur between two traditional Landau phases, also called phases with broken symmetry. 
Furthermore, there is an ongoing intense theoretical effort in identifying and understanding
topological phase transitions, i.e. quantum phase transitions between two topologically distinct phases.
Thus it appears that in many cases
quantum phase transitions occur between phases belonging to the same class,
i.e. either between two topological or two traditional Landau phases.
In contrast, the transition between the paired FQHS and the QHN
is a rare transition that occurs across the two distinct classes of phases, i.e.
between a topological and a traditional Landau phase.
Near the quantum critical point of this transition
the vanishing topological order is accompanied by the emergence of a broken symmetry. 

The 2DEG may support other potential examples
of transitions from a topological to a traditional Landau phase, such as that from
the terminal FQHS at the highest magnetic fields
and the Wigner crystal \cite{rip1,rip2}. However, for the phase associated with the Wigner crystal,
translational order has not yet been demonstrated; the Wigner crystal is identified
from an insulating behavior in transport accompanied by the observation of pinning resonances
in the microwave frequency domain \cite{rip3,rip4}. Pinning resonances are strong
deep in the insulating regime, leaving the possibility open for an  
Anderson-type of insulator with no translational order close to the transition point.
In contrast, high pressure studies clearly demonstrate broken rotational symmetry,
providing therefore a strong evidence for nematic order.

\begin{figure}[b]  %h
\includegraphics[width=3in]{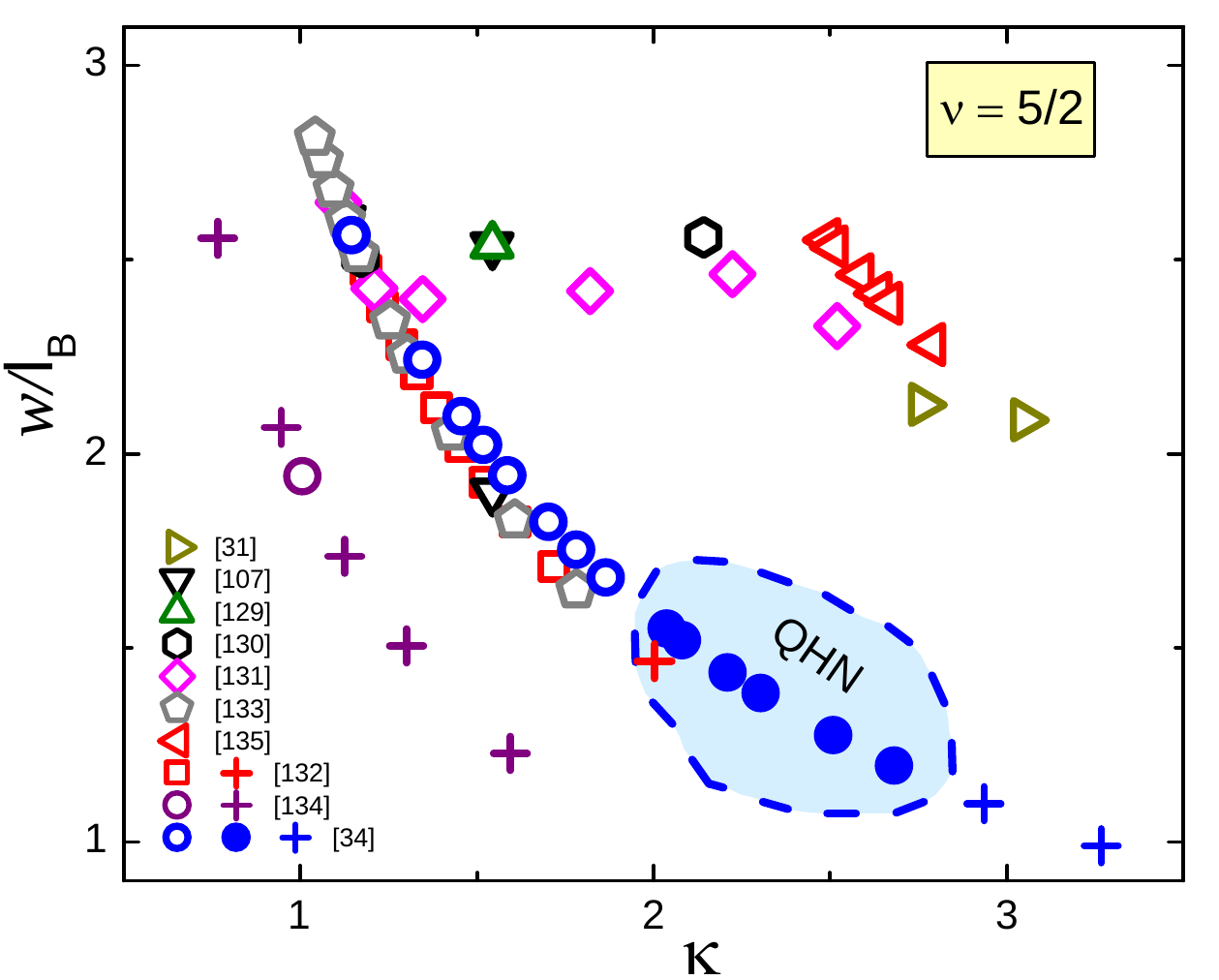}
\caption{
Ground states at $\nu=5/2$ in the $\kappa$-$w/l_B$ plane. Data based on
measurements in the ambient, with the exception of blue symbols, which
are measured under pressure.
Open symbols represent FQHSs, closed ones QHNs. 
For the $+$ symbols neither a FQHS nor a
nematic phase was demonstrated. Adapted from Schreiber et al. \cite{kate3}.
}
\label{fig5}
\end{figure}

\begin{figure*}[t]  %h
\includegraphics[width=5in]{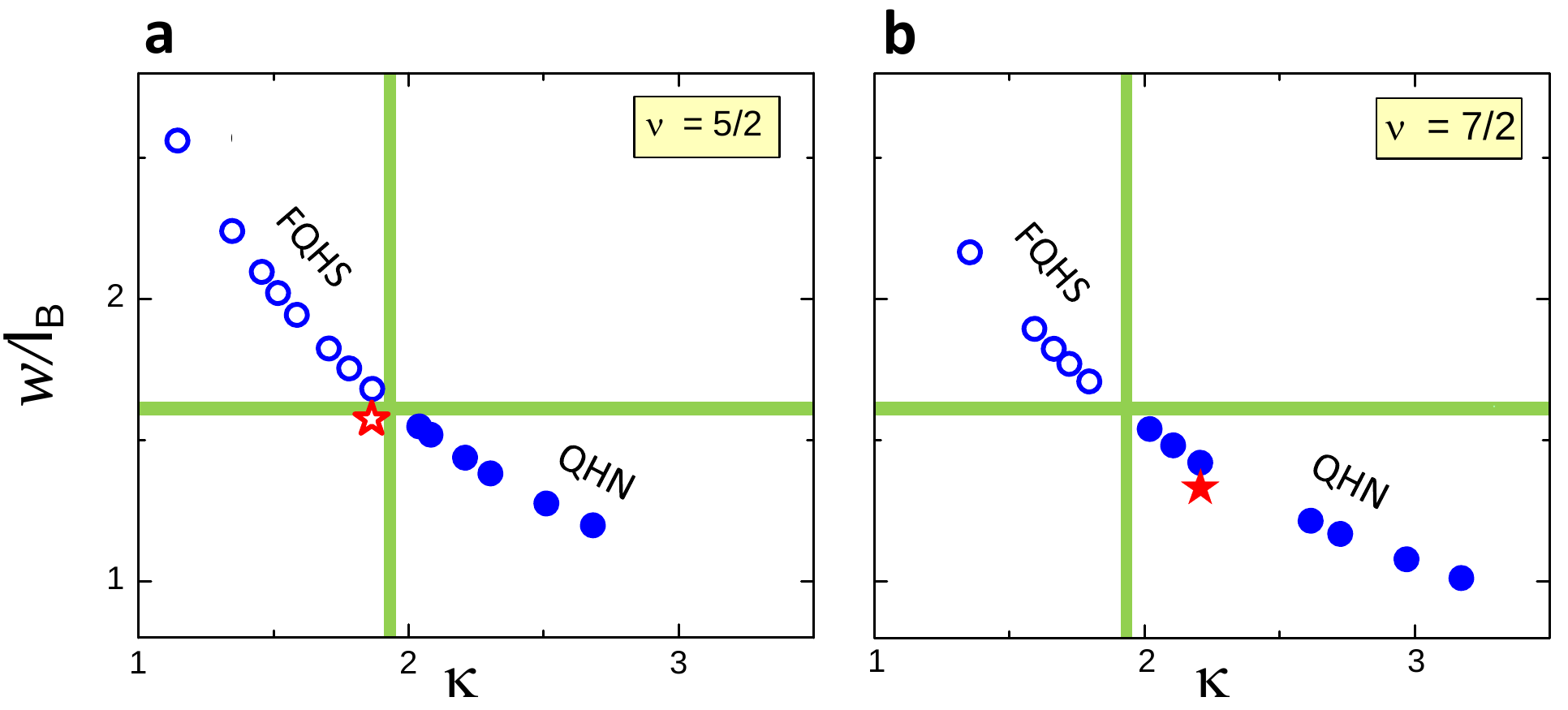}
\caption{Ground states at $\nu=5/2$ (panel (\textbf{a}))
and at $\nu=7/2$ (panel \textbf{b}) of samples from \cite{kate3}
in the $\kappa$-$w/l_B$ plane.
Blue symbols are for sample $A$ under pressure, while
the red symbols are for sample $B$ in the ambient.
Open symbols represent FQHSs, closed ones QHNs. 
Vertical and horizontal lines mark the critical values at the transition point.
Adapted from Schreiber et al. \cite{kate3}.
}
\label{fig6}
\end{figure*}

\subsection{The pressure driven transition at half-filling: an example for competition of pairing and nematicity}

The transition from the FQHS to the QHN was observed at $\nu=5/2$ and $\nu=7/2$,
filling factors at which the FQHS is due to pairing of the composite fermions. 
Of equal importance is the fact that in perpendicular magnetic fields
FQHSs in which pairing does not play a role, did not exhibit a competition with the
QHN. Two such incipient FQHSs are marked at $\nu=7/3$ and $8/3$
in \textbf{Figure \ref{fig3}}. In this figure we observe that
the QHN centered to $\nu=5/2$ indeed does not extend to $\nu=7/3$ or $8/3$. 
Therefore experiments done at high pressures show that
among the FQHSs in the $N=1$ Landau level, in perpendicular magnetic fields 
only the paired FQHSs,
i.e. the ones at $\nu=5/2$ and $7/2$, may compete with the QHN. 

We then conclude that experiments performed at high pressures revealed that
pairing and nematicity are intimately connected in the $N=1$ Landau level of the 2DEG \cite{kate3}.
In light of the proximity of the FQHSs at $\nu=5/2$ and $\nu=7/2$ to nematicity,
it is perhaps not surprising that the QHN can be stabilized at these filling factors.
However, the two phases do not have to be contiguous, i.e.
there is no fundamental reason for a direct quantum phase transition between
the two phases. The existence of the quantum critical point 
is unlikely to be accidental and it highlights a deep connection between pairing and nematicity \cite{kate3}.

\section{IDENTIFYING THE DRIVING FORCE OF THE TRANSITION}

The quantum phase transition from the paired FQHS to the QHN hinges on a
delicate energy balance of these phases near the critical pressure.
An interesting question then is what is the role of the pressure in driving this phase transition. 
Data shown in \textbf{Figure \ref{fig2}}
clearly demonstrate that the critical pressure of the transition from the paired FQHS to the
QHN at $\nu=7/2$ is different, in fact it is significantly lower than that at $\nu=5/2$ \cite{kate3}.
Why are these two critical pressures different, which sample parameters, if any, 
influence the value of the critical pressure, and may similar 
transitions be induced using a parameter other than pressure?

We remind the reader that in the 2DEG in GaAs/AlGaAs 
the electron density decreases with an increasing pressure.
The density at the critical pressure of the paired FQHS to QHN transition
in \cite{kate1} was found $n_c \simeq 1.1 \times 10^{11}$~cm$^{-2}$. 
The $\nu=5/2$ filling factor was studied in samples 
close to such a density, but the nematic phase was not observed at ambient pressures
\cite{tilt3,dean08,nodar11,nuebler10,watson15}.
A parameter which influences the ground state of the 2DEG
and which changes with the density is the Landau level mixing parameter $\kappa$, defined
as the ratio of the Coulomb and cyclotron energies \cite{yoshi}. 
$\kappa$ strongly affects for the $\nu=5/2$ FQHS \cite{sim3,sim4,sim5,sim6,sim7,sim8,sim9}
and the QHN \cite{kennett}.
Since the $\nu=5/2$ QHN was not observed in the ambient at any
studied density, parameters other than $\kappa$ must also play a role
in stabilizing the QHN. As suggested by Rezayi and Haldane \cite{haldane-str}
and in other numerical work \cite{sim2,sim3,sim6,sim7}, one such parameter
is the width of the wavefunction in the direction perpendicular to the plane of the 2DEG.

We think that the phase transition ocurring near the critical pressure is driven by the
electron-electron interaction \cite{kate3}. The role of the electron-electron interaction
in stabilizing different ground states of the 2DEG is well known:
in the most general case, the electron-electron interaction is modified from its Coulomb
expression and it is a function of the orbital Landau level index $N$,
the Landau level mixing parameter $\kappa$, and
the finite thickness of the electron layer in the direction perpendicular
to the plane of the electrons. The latter quantity may be approximated by the
width of the quantum well $w$, which is relevant in its dimensionless form
$w/l_B$. Here $l_B$ is the magnetic length \cite{jainBook}.

Since both $\kappa$ and $w/l_B$ depend on the density, they will also 
change with pressure. The usefulness of these parameters becomes evident
when we mark the different ground states at $\nu=5/2$ in the $\kappa$-$w/l_B$ parameter
space.  \textbf{Figure \ref{fig5}} contains data obtained in the ambient
\cite{tilt3,dean08,nodar11,pan12,nuebler10,watson15,shi15,nodar17} and also
results from our high pressure experiment \cite{kate3}.
Data from Refs.\cite{nuebler10,watson15,shi15,nodar17} are collected 
from density-tunable samples. Since at a constant band mass,
dielectric constant, and at a fixed filling factor $w/l_B \propto 1/\kappa$,
these data points follow a hyperbola in the $\kappa$-$w/l_B$ space.
The curve from Ref.\cite{kate3} deviates slightly from such a dependence due to the
variation with pressure of the band mass and dielectric constant \cite{p1}.
We notice that the QHN is in the lower left area of this plot, in
a region not yet accessed with 2DEGs in the ambient.
This finding provides a natural explanation why the QHN was not
observed at $\nu=5/2$ (that is in purely perpendicular magnetic fields) in prior studies.

Further insight may be gleaned from a comparison of the nematic onset at $\nu=5/2$ and $7/2$.
In  \textbf{Figure \ref{fig6}} the transition from the
paired FQHS to the QHN occurs at the crossing of the two green lines
at the spin index-independent critical value $\kappa_c \simeq 1.9$ and $w/l_{B,c} \simeq 1.6$.
In contrast, the critical pressure of the transition at $\nu=5/2$ 
is significantly different from that at $\nu=7/2$.
The independence of $\kappa_c$ and $w/l_{B,c}$ at the critical point on the spin index
suggests that the transition from the paired FQHS to QHN 
is driven by the electron-electron interaction, as parametrized by the Landau level mixing parameter $\kappa$
and dimensionless width $w/l_B$ of the quantum well \cite{kate3}.

\begin{figure}[t]  %h
\includegraphics[width=3in]{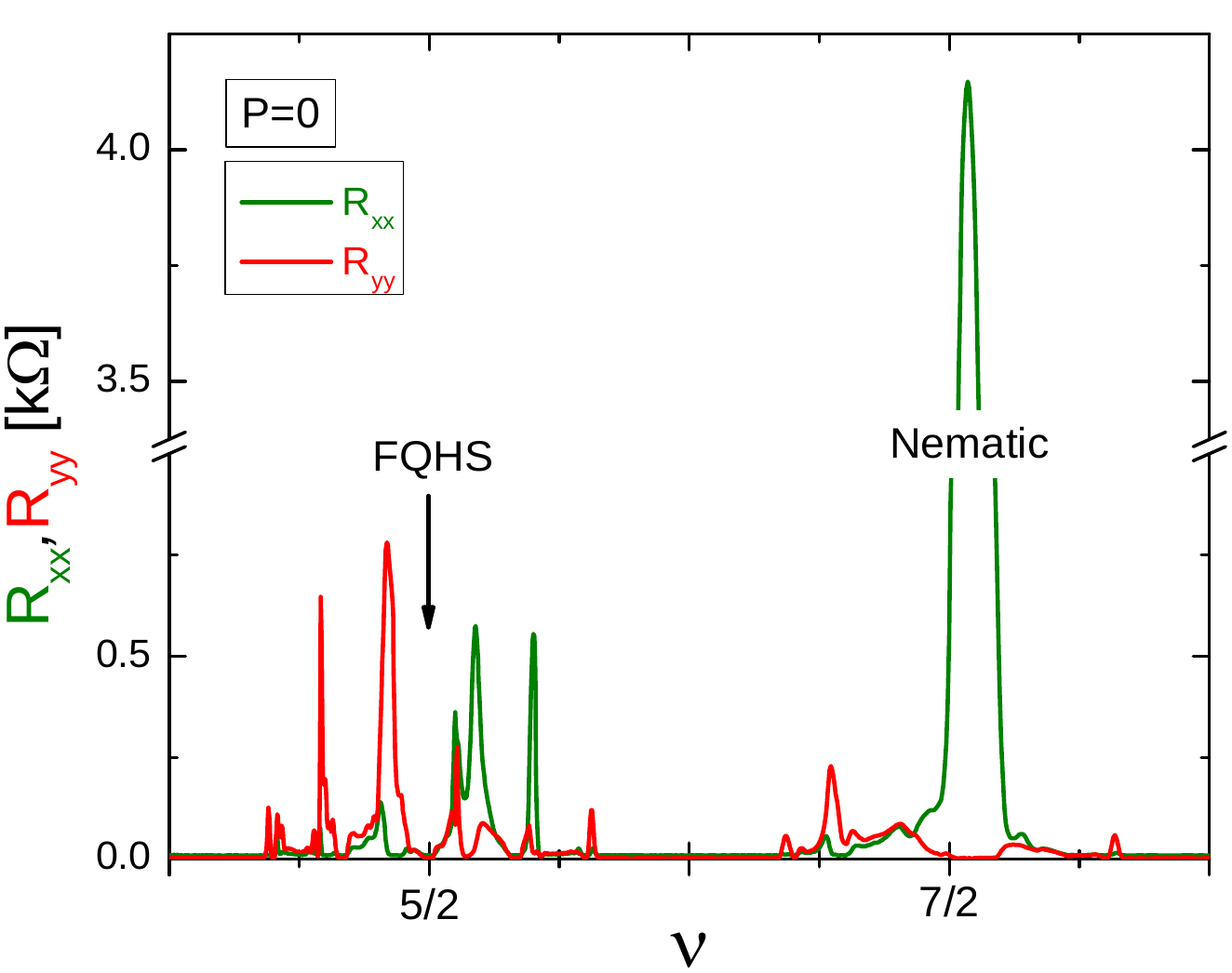}
\caption{
Magnetoresistance in the second Landau level of sample $B$ from \cite{kate3}
measured at ambient pressure and at $T \simeq 4.5$~mK. The
strong resistance anisotropy at $\nu = 7/2$ signals the QHN, while at $\nu = 5/2$ we observe a FQHS.
Adapted from Schreiber et al. \cite{kate3}.
}
\label{fig7}
\end{figure}

As a final test for the relevance
of the electron-electron interaction, we investigated a sample to be measured
at ambient pressure, but in which the electron-electron interaction was 
engineered near its value at the quantum critical point \cite{kate3}. This sample, labeled $B$ in Ref.\cite{kate3},
was also based on a $30$~nm quantum well, but it had a density of 
$n = 1.09 \times 10^{11}$~cm$^{-2}$. These growth parameters
resulted in calculated $\kappa$ and $w/l_B$,
shown by red stars in \textbf{Figure \ref{fig6}}, that fall in the range of the nematic at $\nu=7/2$.
Magnetoresistance traces for
this sample, as measured in a $^3$He immersion cell assuring electron 
thermalization below $5$~mK, are shown in \textbf{Figure \ref{fig7}}. 
At $\nu=7/2$ an extremely large resistance anisotropy was indeed observed, signaling a
strong QHN. 
Furthermore, at $\nu=5/2$ a weak FQHS was seen, consistent with the $\kappa$ and 
$w/l_B$ parameters being in near proximity to the critical values at the transition.
Taken together, there is compelling evidence that the QHN
is stabilized at $N=1$ at ambient pressure
when the electron-electron interaction is tuned via the parameters
$\kappa$ and $w/l_B$ to the stability range of the nematic. Furthermore, data suggest
that a transition to the nematic can also be induced at ambient pressure,
by tuning the electron-electron interaction via the density.

\section{OUTLOOK AND OPEN QUESTIONS}

We surveyed various aspects of the competition of pairing and nematicity
in the $N=1$ Landau level of the 2DEG in GaAs/AlGaAs. The interplay of related
orders also occurs in unconventional superconductors. However, in contrast to the latter
systems, the 2DEG is simpler in many respects and it therefore may offer
insight. Indeed, the 2DEG is a single band system, both pairing and nematicity are known to be
orbitally driven, and it is a platform with highly tunable 
electron-electron interaction. Perhaps the most surprising finding is 
the deep connection between pairing and nematicity as evident from
the direct quantum phase transition between the paired FQHSs and the QHN.
We showed that this transition is induced by a delicate tuning of the electron-electron interaction.

The origins of the transition from the paired FQHS to the QHN remain elusive.
We argued that this transition is beyond Landau's paradigm.
Similar phase transitions were found in recent models based on
either a quadrupolar interaction between the electrons \cite{th1}
or a built-in mass anisotropy \cite{th2}. However, such effects do not seem to be present
in the 2DEG in GaAs/AlGaAs. What is certain that fluctuations
both in the nematic order parameter and in the Chern-Simons gauge field
contribute to the destruction of order near the quantum critical point \cite{th1,th3}.
Fluctuation effects remain experimentally unexplored.

The existence of the quantum critical point has implications not only for the ordered phases, but also
for the parent Fermi liquid from which they condense \cite{anders}.
The FQHS forming below the critical pressure cannot be accounted for without
the formation of the composite fermions. In contrast,  composite fermions do not
have to be invoked for the description of the QHN.
In the $N=1$ Landau level there is therefore a dichotomy of two descriptions:
one based on composite fermions and another on electrons. This dichotomy inevitably raises the question
whether or not it also applies to the parent Fermi liquid \cite{kate2}.
The Fermi liquid near the quantum critical point is a candidate for a
strange metal which may exhibit a crossover or a transition
from a Fermi liquid of composite fermions to a Fermi liquid of electrons.
This strange metal is likely related to the non-Fermi liquid proposed in recent theories
group calculation \cite{th1,th3}. 
The existence of such a strange metal may provide yet another link between
the physics of the 2DEG at half filling and that of unconventional superconductors \cite{anders}.

There are numerous avenues for future experimental work.
Even though it is expected, the quantum
Hall nematic at $\nu=5/2$ has not yet been seen at ambient pressure.
The search for novel phases in the quantum critical region, such as the recently 
proposed pair-density-wave state \cite{th6}, may be fruitful. 
A more thorough mapping of the $\kappa$-$w/l_B$ space is needed and
studies under uniaxial strain and tilt will also likely to offer new insight.
The evolution of the reentrant integer quantum Hall states toward 
the quantum critical point may be interesting. Currently it is
not known whether the QHN occurs in other 
high mobility 2DEGs. This question may be especially relevant for high
quality materials exhibiting paired FQHSs, such as ZnO/MgZnO \cite{zno},
bilayer graphene \cite{gr0,gr1,gr2}, and monolayer graphene \cite{gr3,gr4}.
The recent discovery of reentrance of the IQHS in graphene, a phase associated
with the electronic bubble phase closely related to the QHN, provided a first hint
that complex charge order is possible in a clean host other than GaAs/AlGaAs \cite{gr-riqh}.

Finally, we argued that pressure is not a primary driving parameter
in stabilizing the QHN.
Nonetheless, pressure still plays a subtle role. According to \textbf{Figure \ref{fig6}b},
the point associated with $\nu=7/2$ in sample $B$ from \cite{kate3} measured in the ambient
is deep in the nematic region. However, its nematic onset is at $T_{onset}^{7/2} \simeq 9.5$~mK,
far lower than $T_{onset}^{7/2} \simeq 42$~mK, the expected value
in the pressurized sample $A$ from Ref.\cite{kate3}at the same density.
The significantly lower nematic onset
in the sample in the ambient is currently not understood.

% Acknowledgements
\section*{ACKNOWLEDGMENTS}

This work was supported by the DOE BES award DE-SC0006671.
We thank our collaborators Nodar Samkharadze for his work on the described projects, 
Michael Manfra, Loren Pfeiffer, and Ken West for the GaAs/AlGaAs for providing samples of exquisite quality, and 
Rudro Biswas, Eduardo Fradkin, and Yuli Lyanda-Geller for help with understanding the data.
We also acknowledge numerous  illuminating discussions with
James Eisenstein, Jainendra Jain, Edward Rezayi, and Boris Shklovskii.

%\section*{LITERATURE\ CITED}

%\\

%\noindent

\end{document}